Title: Further metrics for flash-radiography source spots

Author(s): Ekdahl, Carl August Jr.

# Further metrics for flash-radiography source spots

Carl Ekdahl[1,*]

[1]*Los Alamos National Laboratory, PO Box 1663, Los Alamos NM 87545*

*Corresponding author: cekdahl@lanl.gov*

Flash radiography of large hydrodynamic experiments driven by high explosives is a diagnostic method in use at many laboratories. A major contributor to the practical resolution obtained by this point-projection technique is the size of the source spot. A variety of techniques for measuring spot size are in use at the different laboratories, as are different definitions of spot size. Additional metrics in common use at Los Alamos National Laboratory are considered here, and compared with those previously discussed in this journal.

OCIS codes: 340.7440, 350.6980, 110.3000, 110.4100

## 1. INTRODUCTION

Flash radiography of large hydrodynamic experiments driven by high explosives is used as a diagnostic tool at many laboratories. Typically multi-MeV radiation pulses with tens of ns pulse widths are used to radiograph metals driven by high explosives in order to obtain a stopped-motion image [1,2,3]. Most often the source of radiation is bremsstrahlung from electrons striking a heavy-metal target, and there has been a substantial investment in high-power electron accelerators for this purpose [2,3]. This is point-projection radiography, in which the object is back lit with a small source of penetrating radiation. Resolution is ultimately limited by the size of the bremsstrahlung source spot, and a variety of techniques for measuring spot size has evolved at the different laboratories, along with many "spot size" definitions [4-7]. These definitions all depend on the distribution of radiation over the spot, which can be vastly different for different machines.

In an earlier article in this journal [7], the metrics that were considered were the full width at half maximum (FWHM), the Los Alamos definition of spot size ( $d_{LANL}$ ), the AWE definition of spot size ( $d_{AWE}$ ), and the limiting resolution (LR). These are defined in the Appendix. Since that earlier article it has become apparent that two important metrics should be added. The first of these is the rms beam (or spot) radius, which is commonly used by accelerator physicists. The second is the FWHM of the line spread function (LSF), which is commonly reported by data analysts. (The LSF is also defined in the Appendix.)

We apply these metrics to some examples that can be treated analytically. These are listed in Table I, and were chosen to cover the range of central tendencies that one encounters in real experimental data, from broad distributions to sharply peaked ones with large wings. Three of these (uniform spot, Gaussian, and Bennett) have been used in the past to describe experimental data [4-7]. We also included the quasi-Bennett approximation to the sum of exponentials used to describe the LSF of the spot produced by the first accelerator at the Los Alamos Dual Axis Radiographic Hydrodynamic Testing (DARHT) facility [10]. Table II compares all of the different metrics for the example distributions. Since the rms radius is ill-defined for both the Bennett and Quasi-Bennett profiles, it is tabulated as na (not applicable).

## 2. APPLICATION

We offer two practical examples of how these results are used. At the DARHT facility the results of pinhole spot-size measurements are routinely reported in real-time both as $d_{LANL}$ and FWHM of the LSFs in the $x$ and $y$ directions. Simple ratios of these can provide immediate information about the shape of the distribution without resorting a full moment analysis (e.g. calculation of the distribution skewness and kurtosis). For example, the ratio of $d_{LANL}$ to FWHM(LSF) is 1.2 for a Uniform Disk, 1.6 for a Gaussian, and 2.7 for a Bennett. That is, a ratio less than 1.6 is indicative of a platykurtic distribution, while a ratio greater than 1.6 is indicitive of a leptokurtic distribution. This is be useful information for operation of the accelerator, because a highly leptokurtic distribution with "wings" is indicative of electron-beam halo that might be reduced by retuning.

The second example is that results of calculations of spot size based on accelerator physics are usually given as the rms radius, or the equivalent envelope radius, which is $\sqrt{2}$ time the rms radius. On the other hand, the radiographers report $d_{LANL}$ from actual measurements, so relating these two metrics facilitates succinct communications. For example, spot sizes that are typically nearly Gaussian have $d_{LANL}$ equal to ~2.7 times the rms radius, so if a beam dynamics simulation predicts a 1-mm rms spot radius, one might expect the radiographers to report 2.7 mm $d_{LANL}$ from actual measurements.

## ACKNOWLEDGEMENTS

The author wishes to acknowledge stimulating discussions with Tom Beery, Evan Rose, B. Trent McCuistian, and Scott Watson on these, and other, topics. This work was supported by the National Nuclear Security Agency of the United States Department of Energy under contract number DE-AC52-06NA25396.

## APPENDIX A. Definitions

### A.1. Point Spread Function (PSF)

An imaging system can be characterized by a point-spread function (PSF), which is normalized so that

$$\int_{-\infty}^{\infty}\int_{-\infty}^{\infty} \mathrm{PSF}(x, y)\, dxdy = 1 \qquad (1)$$

or

$$\int_{0}^{\infty} \mathrm{PSF}(r) 2\pi r dr = 1 \qquad (2)$$

for the symmetric intensity distributions considered here.

For point-projection flash radiography with the resolution limited by the source spot, rather than the imaging optics, the point-spread function is just the normalized intensity distribution of the spot,

$$\mathrm{PSF}(x, y) = I(x, y) / I_{TOTAL}\,. \qquad (3)$$

Experimentally, the point spread function is the result of a pinhole measurement of the spot after de-blurring by deconvolving the pinhole PSF.

### A.2. Line Spread Function (LSF)

The line-spread function (LSF) is the projection of the PSF onto a plane; for example

$$\mathrm{LSF}(x) = \int_{-\infty}^{+\infty} \mathrm{PSF}(x, y)\, dy \qquad (4)$$

The LSF can also be computed by using the forward Abel transform

$$\mathrm{LSF}(x) = 2\int_x^{\infty} \frac{r\,\mathrm{PSF}(r)}{\sqrt{r^2 - x^2}}\,dr \qquad (5)$$

Conversely, if the LSF is known, the PSF can be found from the inverse Abel transform

$$\mathrm{PSF}(r) = -\frac{1}{\pi}\int_r^{\infty} \frac{d\mathrm{LSF}/dx}{\sqrt{x^2 - r^2}}\,dx \qquad (6)$$

### A.3. Edge Spread Function (ESF)

The edge-spread function is the integral of the LSF

$$\mathrm{ESF}(x) = \int_{-\infty}^{x} \mathrm{LSF}(x')\,dx' \qquad (7)$$

Experimentally, the edge spread function is directly produced by an ideal roll-bar measurement of the spot.

### A.4. Modulation Transfer Function (MTF)

The modulation transfer function (MTF) is defined as the modulus of the 2-dimensional Fourier transform of the point-spread function of the x-ray imaging system;

$$\mathrm{MTF}(\nu_x,\nu_y) = \left|H(\nu_x,\nu_y)\right| \quad , \qquad (8)$$

where,

$$\mathrm{H}(\nu_x,\nu_y) = \int_{-\infty}^{\infty} dx \int_{-\infty}^{\infty} \mathrm{PSF}(x,y)\exp\left[-2\pi j(\nu_x x + \nu_y y)\right]dy \ . \qquad (9)$$

Here, $k_i = 2\pi\nu_i$ is the wavenumber in the particular direction, so $\nu_i$ is often called the spatial frequency with units of inverse dimension. Note that the MTF is a two dimensional surface in wavenumber space.

A useful analytic tool is the projection-slice theorem, which states that the Fourier transform of the projection (LSF) yields a slice of the modulation transfer function orthogonal to the projection. That is, a slice of the two-dimensional MTF can be calculated by projecting the PSF onto a plane to produce the Line Spread Function (LSF), and then taking the one-dimensional Fourier transform of the LSF. This can be verified by inspection of Eq. (9) by setting one of the spatial frequencies to zero. For example, consider the slice of the MTF in the $k_x$ direction, which is given by

$$\mathrm{MTF}(\nu_x, 0) = \left|\mathrm{H}(\nu_x, 0)\right| = \left|\int_{-\infty}^{\infty} dx \int_{-\infty}^{\infty} \mathrm{PSF}(x, y) \exp\left[-2\pi j(\nu_x x)\right] dy\right| \tag{10}$$

It is worth noting that the MTF can be found directly from a known (measured) line spread function by taking its Fourier transform. The integral over $y$ in Eq. (10) is just the line spread function LSF($x$), according to the definition Eq. (4), so it follows that

$$\mathrm{MTF}(\nu_x, 0) = \left|\int_{-\infty}^{\infty} \mathrm{LSF}(x) \exp\left[-2\pi j(\nu_x x)\right] dx\right| \tag{11}$$

### A.5. Full-Width at Half Maximum

Quite frequently, the spot size is characterized by its width at 50% of peak value, which is known as the full-width at Half Maximum (FWHM). A related, infrequently quoted metric is the full width at some other percentage of the peak value. For example, the 10% width is sometimes used.

### A.6. AWE Spot Size

This metric was developed at the Atomic Weapons Establishment (AWE) as a means for rapid characterization data. The AWE spot size is defined in terms of the horizontal separation $\Delta_x$ between two values of the measured ESF. This distance is then multiplied by a number that would have given the same separation for a uniformly filled disk of diameter $d_{AWE}$ [6-10]. Traditionally the two values of the ESF are chosen to be 25% and 75% of the maximum. For this choice the multiplier is 2.47541, and the AWE spot size is

$$d_{AWE} = 2.47541\Delta_x \tag{12}$$

This is frequently rounded off to $d_{AWE} = 2.5\Delta_x$.

**A.7. Los Alamos Spot Size**

In flash radiography at Los Alamos National Laboratory (LANL), the source spot size is frequently characterized by its "50%MTF size." Simply stated, the LANL spot size, $d_{LANL}$, is the diameter of a uniformly illuminated disk that has the same spatial frequency at MTF=0.5 as the actual source spot[4,5]. In practice, one computes the MTF of the source spot, finds the spatial frequency at which the MTF = 0.5, $\nu_{1/2} = k_{1/2} / 2\pi$ , and divides this into 0.70508 to get the equivalent disk diameter. That is,

$$d_{LANL} = 0.70508 / \nu_{1/2} = 4.4301 / k_{1/2} \tag{13}$$

**A.8. Limiting Resolution**

Using the MTF, the limiting resolution prescription is straightforward: from measurements or theory simply calculate the MTF, and then find the spatial frequency at which the MTF is 5%. That is, $\mathrm{LR} \equiv \nu_{LR} = k_{LR} / 2\pi$ (lp/mm), where $\mathrm{MTF}\left(k_{LR}\right) = 0.05$. This is equivalent to the separation of line pairs in an Air Force Target that are barely distinguishable by normal human vision.

## A.9. RMS size

The rms beam size is defined as the second moment of the LSF. For example, the $y_{rms} = \left\langle y^2 \right\rangle^{1/2}$ where

$$\left\langle y^2 \right\rangle = \int_{-\infty}^{\infty}\int_{-\infty}^{\infty} y^2 \, \mathrm{PSF}(x, y)\, dxdy \tag{14}$$

For the symmetric intensity distributions considered here, one has

$$\left\langle r^2 \right\rangle = \int_0^{\infty} r^2 \mathrm{PSF}(r) 2\pi r dr \tag{15}$$

In general $\left\langle r^2 \right\rangle = \left\langle x^2 \right\rangle + \left\langle y^2 \right\rangle$ and since $x_{rms} = y_{rms}$ for a symmetric distribution, one has $r_{rms} = \sqrt{2}\, y_{rms}$ .

Table I. Example source distributions

| Source Distribution | PSF($r$) | LSF(x) | MTF($k$) |
|---|---|---|---|
| Uniform Disk | $\frac{1}{\pi a^2} \quad r \le a$<br>$0 \quad r > a$ | $\frac{2}{\pi a^2}\sqrt{a^2 - x^2} \quad x < a$<br>$0 \quad x \ge a$ | $2\frac{\lvert J_1(ka)\rvert}{ka}$ |
| Gaussian | $\frac{1}{\pi a^2}\exp\left(-r^2 / a^2\right)$ | $\frac{1}{a\sqrt{\pi}}\exp\left(-x^2 / a^2\right)$ | $\exp\left(-k^2 a^2 / 4\right)$ |
| Bennett | $\frac{1}{\pi a^2}\frac{1}{\left[1+\left(r / a\right)^2\right]^2}$ | $\frac{a^2}{2\left(a^2 + x^2\right)^{3/2}}$ | $ka\ K_1(ka)$ |
| Quasi-Bennett | $\frac{1}{2\pi a^2}\frac{1}{\left[1+(r / a)^2\right]^{3/2}}$ | $\frac{a}{\pi\left(a^2 + x^2\right)}$ | $e^{-ka}$ |

Table II. Spot size metrics for the example source distributions.

| Source Distribution | FWHM | FWHM (LSF) | $d_{LANL}$ | $d_{AWE}$ | $r_{rms}$ | LR |
|---|---|---|---|---|---|---|
| Uniform Disk | $2a$ | $1.732a$ | $2a$ | $2a$ | $0.707a$ | na |
| Gaussian | $1.665a$ | $1.665a$ | $2.661a$ | $2.082a$ | $a$ | $0.551/a$ |
| Bennett | $1.287a$ | $1.533a$ | $3.474a$ | $2.858a$ | na | $0.636/a$ |
| Quasi-Bennett | $1.533a$ | $2.000a$ | $6.351a$ | $4.951a$ | na | $0.477/a$ |